\documentclass[useAMS,usenatbib]{mn2e}
\usepackage{amssymb}
\usepackage{graphics}
\usepackage{multicol}

\def\eqref#1{equation~(\ref{#1})}

\def\apref#1{Appendix~\ref{#1}}

\def\binom#1#2{{#1 \choose #2}}

\title[Properties of analytic transit light curve models]{Properties of analytic transit light curve models}
\author[A. P\'al]{%
Andr\'as P\'al$^{1,2}$\thanks{E-mail: apal@cfa.harvard.edu
}\\
$^{1}$Harvard-Smithsonian Center for Astrophysics,
        60 Garden street,
        Cambridge, MA, 02138, USA \\
$^{2}$Department of Astronomy, Lor\'and E\"otv\"os University,
        P\'azm\'any P. st. 1/A,
        Budapest H-1117, Hungary}

\begin{document}

\date{Accepted \dots. Received \dots; in original form \dots}

\pagerange{\pageref{firstpage}--\pageref{lastpage}} \pubyear{2008}

\maketitle

\label{firstpage}

\begin{abstract}
In this paper a set of analytic formulae are presented with which 
the partial derivatives of the flux obscuration function 
can be evaluated -- for planetary transits and eclipsing binaries -- 
under the assumption of quadratic limb darkening.
The knowledge of these partial derivatives is crucial for many 
of the data modeling algorithms and estimates of
the light curve variations directly from the changes in the orbital elements.
These derivatives can also be utilized to speed up some of the fitting
methods. A gain of $\sim 8$ in computing time 
can be achieved in the implementation 
of the Levenberg-Marquardt algorithm, relative to using numerical
derivatives.
\end{abstract}



\section{Introduction}
\label{sec:introduction}

In recent years, the discovery and further characterization
of transiting extrasolar planets (TEPs) has provided unique
information about the nature of planetary systems. The analysis of a planet
which periodically eclipses its host star yields the physical 
radius, the inclination and the mass of the system in addition to the parameters
which are gathered from the radial velocity measurements.
Since the discovery of the first such
system \citep[see][]{charbonneau2000,brown2001}, more than 40 
other TEPs were discovered around other stars. Currently operating
ground-based surveys are producing numerous new discoveries.
The doubling period of the number of known TEPs is below one
year. Moreover, existing \citep[CoRoT, see][]{barge2008,alonso2008} and 
planned space-borne instruments
\citep[e.g. Kepler Mission, see][]{borucki2007}
are expected to yield hundreds of new discoveries of such systems, 
even with planetary radii comparable to that of Earth. 
Also, subsequent observations
of a given transiting system can provide some information on
the variations in the timing of successive transits 
and the light curve shape.
These detections can be used to constrain other planetary 
companions \citep{agol2005,steffen2007,holman2005} or co-orbital companions
\citep[Trojans, see][]{ford2007}.

In order to optimize the precision and speed of TEP observations, a
careful analysis of the light curves is required. The basis of such light
curve analysis is to find an adequate model of planetary obscuration,
which causes a small decrease in the stellar flux. 
Since both the stellar and the planetary body can be
well modelled by a spheroidal shape, the decrease in the stellar flux
can be estimated from the full or partial overlap of two circles. The star
itself has a non-negligible limb darkening which depends on both
the stellar properties and the photometric band 
and quantified by a small set of coefficients
\citep[see e.g.][for such tables of limb darkening constants]{claret2004}.
At present, the most widely used models for this problem have been 
given by \citet{mandel2002} 
and \citet{gimenez2006}. \citet{mandel2002} calculate closed form expressions
for the flux decrease assuming non-linear or quadratic limb darkening
(quantified by 4 or 2 coefficients, respectively) 
while \citet{gimenez2006} gives an infinite
series where the limb darkening can be taken into account up
to arbitrary order. 
In most cases the quadratic case is adequate
because the photometric precision of typical data is not 
good enough for the higher 
order limb darkening models to make a difference.

Most data modeling algorithms, including the well known 
nonlinear Levenberg-Marquardt fitting method \citep[see][]{press1992} 
utilize the partial derivatives of the model function with respect
to the model parameters. 
The uncertainties in the model parameters can be well characterized by
the Fisher information matrix \citep[see e.g.][]{finn1992},
also requiring knowledge of the same partial derivatives.
Therefore, in the case of planetary transits, the parametric 
derivatives of the flux decrease function can be extremely valuable.
Moreover, partial derivatives can be used 
to construct a set of uncorrelated parameters of the light curve
which is preferred by most of the parameter fitting algorithms
(e.g. Levenberg-Marquardt, downhill simplex, Markov Chain Monte Carlo).
Also, the analysis of the partial derivatives with respect to the 
limb darkening coefficients themselves yields a combination of 
these with which consistent sanity checks can be done verifying
the stellar atmospheric properties in an independent way. 
Finally, these derivatives can be used to directly calculate
how the variations in the orbital parameters
affect transit timing and the shape of the light curve.

In this paper, we present the partial derivatives of the
flux decrease function assuming a quadratic limb darkening law. In the
next section, the formalism and the derivatives are presented.
In Section 3, we apply these derivatives to construct a set of well behaved
parameterizations of transiting light curves which yield an always finite
and moderate correlation between the adjusted quantities in all important
cases and limits. The correlations
between the limb darkening coefficients are also discussed. 
The results are summarized in the last section.


\begin{table}
\caption{Exclusion cases of different occultation geometries.
To figure out the respective case for a certain value of $p$ and $z$,
check the relations step by step: if the current relation is true, that
case can be assigned to the given values of $p$ and $z$, 
otherwise go to the next step. The final case $\mathbf{G}$
is when there is no obscuration. Cases with subscripts can only occur
if the radius of the planet is not smaller than $1/2$.}
\label{table:cases}
\begin{center}\begin{tabular}{rllr}
\hline
\hline
Step & Relation & Case & M\&A case \\
\hline
1  &	$z=0$ ~\&~ $p<1$	&	$\mathbf{A}$	& 10	\\
2  &	$z\le p-1$		&	$\mathbf{A_G}$	& 11	\\
3  &	$z<p$ ~\&~ $z<1-p$	&	$\mathbf{B}$	& 9	\\
4  &	$z<p$ ~\&~ $z=1-p$	&	$\mathbf{B_T}$	& --	\\
5  &	$z<p$ 			&	$\mathbf{B_G}$	& 8	\\
6  &	$z=p$ ~\&~ $z<1-p$	&	$\mathbf{C}$	& 5	\\
7  &	$z=p=1/2$		&	$\mathbf{C_T}$	& 6	\\
8  &	$z=p$ 			&	$\mathbf{C_G}$	& 7 	\\
9  &	$z<1-p$			&	$\mathbf{D}$	& 3	\\
10 &	$z=1-p$			&	$\mathbf{E}$	& 4	\\
11 &	$z<1+p$			&	$\mathbf{F}$	& 2 	\\
12 &	--			&	$\mathbf{G}$	& 1	\\
\hline
\end{tabular}\end{center}
\end{table}


\section{Parametric derivatives of the flux decrease}
\label{sec:paramderiv}

In this section the partial derivatives of the flux decrease are presented.
The surface brightness of a star as a function of the normalized
distance $0\le r\le 1$ assuming quadratic limb darkening is given
by the equation
\begin{equation}
I(r)=1-\sum\limits_{m=1,2}\gamma_m\left(1-\sqrt{1-r^2}\right)^m,
\end{equation}
where the constants $\gamma_1$ and $\gamma_2$ quantify the limb darkening.
Recalling \citet{mandel2002}, the relative apparent flux of an eclipsed star
with a quadratic limb darkening can be written as $f=1-\Delta f$
(assuming a unity flux out of the transit), where flux decrease $\Delta f$ 
can be calculated using the equation
\begin{eqnarray}
\Delta f & = & 
 	W_0F_0+W_2F_2+ \label{eq:fluxdecrease} \\
& &	W_1[F_1+F_K\mathrm{K}(k)+F_E\mathrm{E}(k)+F_{\Pi}\Pi(n,k)]. \nonumber
\end{eqnarray}
In this equation the quantities $W_i$ ($i=0,1,2$) are only functions of
the limb darkening coefficients, namely
\begin{eqnarray}
W_0 & = & \frac{6-6\gamma_1-12\gamma_2}{W}, \\
W_1 & = & \frac{6\gamma_1+12\gamma_2}{W},\\
W_2 & = & \frac{6\gamma_2}{W},
\end{eqnarray}
where $W=6-2\gamma_1-\gamma_2$. In \eqref{eq:fluxdecrease} the 
terms $F_0$, $F_1$, $F_K$, $F_E$, $F_\Pi$ and $F_2$ are only
functions of the occultation geometry, namely the relative
planetary radius $p\equiv R_p/R_\star$ and the normalized projected distance
$z$ between the center of the star and the center of the planet.
In \eqref{eq:fluxdecrease} the functions $\mathrm{K}(\cdot)$,
$\mathrm{E}(\cdot)$ and $\Pi(\cdot,\cdot)$ denote the complete
elliptic integrals of the first, second and third kind, respectively.
The variation in the occultation geometry yields 12 distinct cases
of obscuration, which are summarized in Table~\ref{table:cases}. This
table is an \emph{exclusion} table and should be interpreted as follows.
For a given value of $(p,z)$, the first relation (in step 1) is
checked. If it is true, the appropriate case can be assigned to the
geometry, otherwise the next relation should be checked and so on. 
The different cases are denoted by bold capitals for planetary radii
smaller than $1/2$ and capitals with a subscript which can only 
occur if the radius of the planet is greater than or equal to $1/2$
In practice it would barely occur for planetary companions for earlier
types of stars but it might happen in the cases when a Jovian 
planet transits a later main sequence star (M dwarf).
For the actually most common planet-like applications ($p<1/2$), 
the 7 major cases ($\mathbf{A}$,\dots,$\mathbf{G}$)
are in the order of growing distance between the geometrical centers of 
the planet and the star. In \eqref{eq:fluxdecrease} the expressions
for the terms $F_i$ ($i=0,1,K,E,\Pi,2$), $k$ and $n$ can be 
found in tables \ref{table:fdaux} and \ref{table:fdcoeffs} in \apref{app:a};
after the appropriate geometrical case has been obtained.
We should note here that \eqref{eq:fluxdecrease} is completely equivalent
with the equation found in \citet{mandel2002}, in the first line of
the second paragraph in their Section 4 at page L173. However, this
expansion of \eqref{eq:fluxdecrease} clearly separates the terms which 
depend only on the limb darkening coefficients ($W_m$) 
and the terms which depend only on the occultation geometry ($F_i$). 

\subsection{Partial derivatives with respect to the limb darkening coefficients}

Since in \eqref{eq:fluxdecrease} the only quantities which depend
on the limb darkening coefficients are $W_0$, $W_1$ and $W_2$, the 
partial derivatives of $\Delta f$ with respect to 
$\gamma_m$ ($m=1,2$) can easily be obtained, namely
\begin{eqnarray}
\frac{\partial\Delta f}{\partial \gamma_m} & = & 
 	\frac{\partial W_0}{\partial\gamma_m}F_0+
	\frac{\partial W_2}{\partial\gamma_m}F_2+ \\
& &	\frac{\partial W_1}{\partial\gamma_m}
	[F_1+F_K\mathrm{K}(k)+F_E\mathrm{E}(k)+F_{\Pi}\Pi(n,k)], \nonumber
\end{eqnarray}
where the appropriate derivatives $\partial W_i/\partial \gamma_m$ 
are the following:
\begin{eqnarray}
\frac{\partial W_0}{\partial\gamma_1} & = & \frac{2W_0-6}{W} \\
\frac{\partial W_0}{\partial\gamma_2} & = & \frac{W_0-12}{W} \\
\frac{\partial W_1}{\partial\gamma_1} & = & \frac{2W_1+6}{W} \\
\frac{\partial W_1}{\partial\gamma_2} & = & \frac{W_1+12}{W} \\
\frac{\partial W_2}{\partial\gamma_1} & = & \frac{2W_2}{W} \\
\frac{\partial W_2}{\partial\gamma_2} & = & \frac{W_2+6}{W}.
\end{eqnarray}

\subsection{Partial derivatives with respect to the geometric parameters}

In \eqref{eq:fluxdecrease}, the terms $F_i$ explicitly depend on the
relative planetary radius $p$ and the normalized distance $z$.
There is also an implicit dependence via the complete elliptic
integrals since their parameters $k$ and $n$ are also functions
of $p$ and $z$. The derivation of these partial derivatives are
quite straightforward for the non-degenerate cases, i.e. for
the cases $\mathbf{B}$, $\mathbf{B_G}$, $\mathbf{D}$ and $\mathbf{F}$
since all of the appearing functions in these domains are analytic.
For the other cases, the partial derivatives can be calculated
as the appropriate limits, namely
\begin{eqnarray}
\partial F_i^{\mathbf{A}}   & \equiv & \lim\limits_{z \to 0}	\partial F_i^{\mathbf{B}}, \\
\partial F_i^{\mathbf{B_T}} & \equiv & \lim\limits_{z \to (1-p)-0} \partial F_i^{\mathbf{B}}, \\
\partial F_i^{\mathbf{C}}   & \equiv & \lim\limits_{z \to p-0}	\partial F_i^{\mathbf{B}} = \lim\limits_{z \to p+0} \partial F_i^{\mathbf{D}}, \\
\partial F_i^{\mathbf{C_T}} & \equiv & \lim\limits_{p \to 1/2}	\partial F_i^{\mathbf{C}}, \\
\partial F_i^{\mathbf{C_G}} & \equiv & \lim\limits_{z \to p+0}	\partial F_i^{\mathbf{F}}, \\
\partial F_i^{\mathbf{E}}   & \equiv & \lim\limits_{z \to (1-p)-0} \partial F_i^{\mathbf{B}}. 
\end{eqnarray}
For $\mathbf{A_G}$ and $\mathbf{G}$, all of the derivatives
are obviously 0, except for the case ($p=1$, $z=0$) when the 
partial derivatives do not exist.

Utilizing the parametric derivatives of the elliptic integrals
(see \apref{app:b}), the final form of the partial derivatives of $\Delta f$
with respect to $z$ and $p$ is
\begin{eqnarray}
\frac{\partial\Delta f}{\partial g} & = & 
 	W_0F_{0,g}+W_1F_{1,g}+W_2F_{2,g}+ \label{eq:pdgfluxdecrease} \\
& &	W_1\mathrm{K}(k)
	\left[F_{K,g}-\frac{(F_K+F_E)k_g}{k}+\frac{F_\Pi n_g}{2n(n-1)}\right]+ \nonumber \\
& &	W_1\mathrm{E}(k)
	\left[F_{E,g}+\frac{F_K k_g}{k(1-k^2)}+\frac{F_E k_g}{k}+\right. \nonumber \\
& & 	+\left.\frac{F_\Pi k k_g}{(k^2-n)(1-k^2)}+\frac{F_\Pi n_g}{2(k^2-n)(n-1)}\right], \nonumber
\end{eqnarray}
where $g$ denotes either $p$ or $z$, the appropriate geometric parameter.
The expressions for $F_{0,g}$, $F_{1,g}$, $F_{K,g}$, $F_{E,g}$,
$F_{2,g}$, $k_g$ and $n_g$ can be figured out for all cases 
using the tables in \apref{app:c}, namely Table~\ref{table:fpartaux}
and Table~\ref{table:fpartcoeffs}.

We note here that the computation of these derivatives are even more
simple and faster than the computation of \eqref{eq:fluxdecrease} since
\eqref{eq:pdgfluxdecrease} lacks the complete elliptic integral of the
third kind for which evaluation requires most of the computing time.


\section{Applications}
\label{sec:applications}

In this section we present three simple applications which utilize 
the partial derivatives of the flux decrease function.
All of these applications assume a transiting planetary system on a circular
orbit with a given semimajor axis (relative to the radius of the star)
$a/R_\star$, an impact parameter, $b\equiv (a/R_\star)\cos i$, 
the planetary companion has a fixed mean motion of $n=2\pi/P$ and the
transit occurs at the instance $E$. The relative radius
of the planet is denoted by $p\equiv R_p/R_\star$.
Therefore, the distance between the center of the stellar and 
planetary disk has a time dependence,
\begin{equation}
z^2(t)=\left(\frac{a}{R_\star}\right)^2\sin^2[n(t-E)]+b^2\cos^2[n(t-E)]. \label{eq:classtrparam}
\end{equation}
From now on we assume that the semimajor axis is relatively large, i.e.
the distance can be approximated by 
\begin{equation}
z^2(t)\cong\left(n\frac{a}{R_\star}\right)^2 (t-E)^2 + b^2.\label{eq:zlinapprox}
\end{equation}
In the following parts of this section, we first calculate the correlations
between the limb darkening coefficients, assuming
the orbital parameters and the relative planetary radius to be known.
In the second part of this section, we construct a set of adjusted
parameters which always yields finite (i.e. definitely smaller than unity)
correlations between them in the cases of non-grazing eclipses.
This is relevant for studies of transiting planets 
since \eqref{eq:zlinapprox} yields a unity
correlation between $a/R_\star$ and $b$ in the limit of $p\to 0$ with
or without limb darkening for all impact parameters. In the last
part of this section we present an analytical calculation how the 
uncertainties in the light curve parameters depend on the photometric 
passbands, assuming a Jupiter-sized planet orbiting a solar-type star.
In all of these cases we use the Fisher matrix method
\citep{finn1992} to obtain the uncertainties and correlations
of the fitted parameters. This method gives the covariance matrix as 
\begin{equation}
\left<\delta a_m \delta a_n\right>=\left(\Gamma^{-1}\right)_{mn},
\end{equation}
where
\begin{equation}
\Gamma_{mn}=\sum_i \frac{\partial_m f(\mathbf{a},t_i)\partial_n f(\mathbf{a},t_i)}{\sigma_i^2}, \label{eq:fisherdef}
\end{equation}
$f(\mathbf{a},t_i)$ is the observed flux at the instance $t_i$,
$\mathbf{a}=(a_1,a_2)=(\gamma_1,\gamma_2)$ or 
$\mathbf{a}=(a_1,a_2,a_3,a_4)=(p,a/R_\star,b,E)$ is the vector of the adjusted parameters
(depending on the actual application)
and $\partial_m\equiv\partial/\partial a_m$. Since we are interested only
in the correlations between the parameters and we can expect uniform
uncertainties in the measurements
and uniform data sampling, therefore $\Gamma$ can
be multiplied by any arbitrary constant and the sum in \eqref{eq:fisherdef} can
be replaced by the integral
\begin{equation}
\Gamma_{mn}\propto\int\limits_{t_1}^{t_2} \partial_m f(\mathbf{a},t)\partial_n f(\mathbf{a},t)\mathrm{d}t.\label{eq:fisherintegral}
\end{equation}
Here $t_1 < E-(na/R_\star)^{-1}(1+p)$ and 
$E+(na/R_\star)^{-1}(1+p) < t_2$, assuring
that the ingress and egress are completely observed, independently
from the impact parameter.

\subsection{Correlations between the limb darkening coefficients}
\label{subsec:corrlimbdark}

\begin{figure}
\resizebox{8cm}{!}{\includegraphics{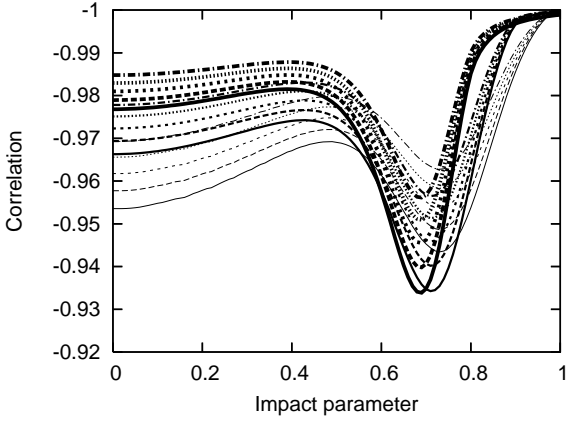}}
\caption{The dependence of the correlation between the
limb darkening coefficients, $C(\gamma_1,\gamma_2)$ as a function
of the impact parameter. Thin lines are for $p=0.01$, moderately
thick lines are for $p=0.1$ and thick lines are for $p=0.2$. The
continuous, long dashed, short dashed, dotted and dotted-dashed lines 
represents the cases when limb darkening coefficients 
are $\gamma_1=\gamma_2=0$, $\gamma_1=\gamma_2=0.1$, $\gamma_1=\gamma_2=0.2$, 
$\gamma_1=\gamma_2=0.3$ and $\gamma_1=\gamma_2=0.4$, respectively.}
\label{fig:ldcorr}
\end{figure}

Now, we determine the correlations between the limb darkening coefficients
when the adjusted parameters are $\mathbf{a}=(\gamma_1,\gamma_2)$. 
It is easy to show that this 
correlation would only depend on the impact parameter, the planetary
radius and the two limb darkening parameters themselves while it does 
not depend on the mean motion, geometrical ratio of $a/R_\star$
and the transit center time $E$. We have obtained these correlations
for very small ($p=0.01$), average ($p=0.1$) and large ($p=0.2$)
planetary radii assuming limb darkening coefficients between $0.0$ and $0.4$,
while the impact parameter was varied between $0$ and $1$. We found that
the correlation is always negative, relatively large, i.e.
$|C(\gamma_1,\gamma_2)|\gtrsim 0.93$ and it strongly depends on
the impact parameter. For these certain values, the correlation
is plotted as a function of $b$ on Fig.~\ref{fig:ldcorr}.
It can easily be seen that the smallest correlation is around 
$b\approx0.7-0.8$, almost independent of the limb darkening 
and the radius of the planet. 

Let us now calculate the optimal linear combination of the limb darkening
coefficients which can be adjusted to yield no correlation. Define 
the parameters $u_1$ and $u_2$ as
\begin{equation}
\binom{u_1}{u_2}=\binom{\cos\varphi~~~~~\sin\varphi}{-\sin\varphi~~\cos\varphi}\binom{\gamma_1}{\gamma_2}.
\end{equation}
For simplicity, let us denote the above orthogonal matrix by $\mathbf{O}=O_{mn}$.
It can be shown that the covariance matrix of $(u_1,u_2)$ and
that of $(\gamma_1,\gamma_2)$ are related to each other as
\begin{equation}
\left<\delta u_k \delta u_\ell\right>=O_{km}\left<\delta \gamma_m \delta \gamma_n\right>\tilde{O}_{n\ell}.
\end{equation}
To make the matrix $\left<\delta u_k\delta u_\ell\right>$ diagonal, the rotation
parameter $\varphi$ should be in coincidence with the 
orientation of the eigenvectors of $\left<\delta \gamma_m \delta \gamma_n\right>$,
namely
\begin{equation}
\varphi = \frac12
\arctan \frac{\left<\delta \gamma_1 \delta \gamma_2\right>+\left<\delta \gamma_2 \delta \gamma_1\right>}	
	     {\left<\delta \gamma_1 \delta \gamma_1\right>-\left<\delta \gamma_2 \delta \gamma_2\right>}.
\end{equation}

\begin{figure}
\resizebox{8cm}{!}{\includegraphics{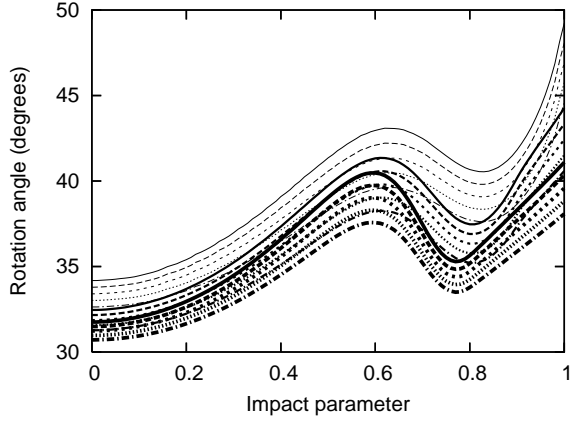}}
\caption{The optimal rotation parameter (in degrees) to avoid correlations
between the limb darkening parameters as a function of the impact parameter
(see text for definition and further details).
Thin lines are for $p=0.01$, moderately
thick lines are for $p=0.1$ and thick lines are for $p=0.2$. The
continuous, long dashed, short dashed, dotted and dotted-dashed lines
represents the cases when limb 
darkening coefficients are $\gamma_1=\gamma_2=0$, $\gamma_1=\gamma_2=0.1$,
$\gamma_1=\gamma_2=0.2$, $\gamma_1=\gamma_2=0.3$
and $\gamma_1=\gamma_2=0.4$, respectively.}
\label{fig:ldoptp}
\end{figure}

We have obtained the optimal values of the rotation parameter $\varphi$
for very small ($p=0.01$), average ($p=0.1$) and large ($p=0.2$)
planetary radii assuming limb darkening coefficients between $0.0$ and $0.4$,
while the impact parameter was varied between $0$ and $1$. We found that
like above, this parameter mostly depends on the impact parameter.
The results are plotted on Fig.~\ref{fig:ldoptp}. The usefulness
of such a plot is somewhat limited if we have no \emph{a priori} knowledge
from the limb darkening coefficients themselves since the correlation
between them  and therefore the optimal rotation
parameter depends on the actual values of $\gamma_1$ and $\gamma_2$.
However, in practice if we have a hint for the planetary radius
and the impact parameter, this angle can be estimated within
a few degrees since the correlation depends more strongly on $b$ and $p$
than $\gamma_1$ or $\gamma_2$. If we do not know any 
reasonable value for $p$ or $b$ before the fit, 
an angle of $\varphi\approx35^\circ-40^\circ$ is a
plausible selection in general.

\subsection{Correlations between the light curve parameters}
\label{subsec:corrlc}

\begin{figure}
\resizebox{8cm}{!}{\includegraphics{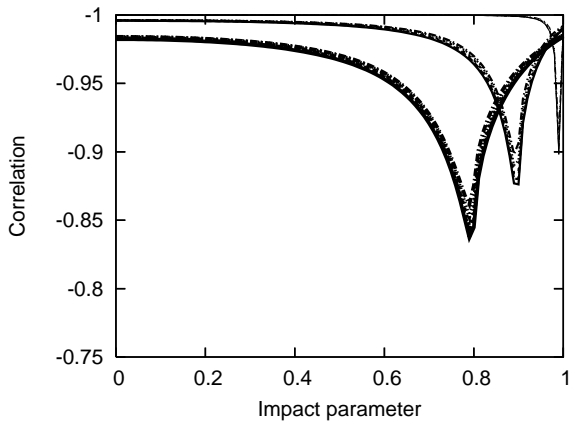}}
\caption{Correlation between the adjusted values of $a/R_\star$ and 
the impact parameter $b$ for various values of planetary radii
and limb darkening coefficients:
Thin lines are for $p=0.01$, moderately
thick lines are for $p=0.1$ and thick lines are for $p=0.2$. The cases
for different limb darkening constants are almost indistinguishable.}
\label{fig:corrarb}
\end{figure}

In this subsection we investigate the correlations between the light
curve parameters utilizing the previously obtained partial derivatives and
Fisher information matrix method. The classical formalism of 
adjusting parameters of a transiting system uses the same
parameters as in \eqref{eq:classtrparam}, namely the ratio $a/R_\star$,
the impact parameter $b$ and the instance of the center of the transit $E$
as well as the radius of the planet, $p=R_p/R_\star$. Since
the flux decrease depends directly on the radius $p$ and indirectly
on $a/R_\star$, $b$ and $E$ (assuming a fixed limb darkening),
the partial derivatives of the light curve $f(t)=1-\Delta f(t)$ 
can be obtained by using \eqref{eq:pdgfluxdecrease} and the chain rule.
These derivatives can then be plugged into \eqref{eq:fisherintegral}
while the adjusted set of parameters will 
be $\mathbf{a}=(p,b,a/R_\star,E)$. We have obtained the correlations
between these variables and we have found that the correlation
between $b$ and $a/R_\star$ tends to unity as the radius of the
planet is decreased. This correlation is plotted as a function
of the impact parameter for planetary radii $p=0.01$, $p=0.1$ and
$p=0.2$ and for various limb darkening parameters on 
Fig.~\ref{fig:corrarb}. It can clearly be seen that $C(b,a/R_\star)$
is almost independent from the actual limb darkening. From 
further analysis of the correlation, it turns out that
for small impact parameters ($b\lesssim 0.5$), $C(b,a/R_\star)$
can be well approximated by $1-p^2/2$. Therefore, for very small
planets, this correlation can be undesirable and most of the 
fitting methods would distort the results. 

\begin{figure}
\resizebox{8cm}{!}{\includegraphics{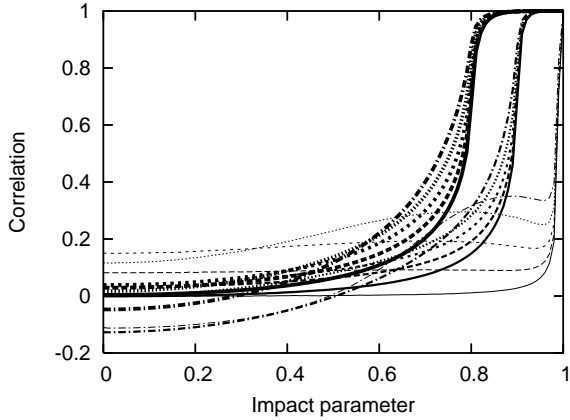}}
\caption{Correlation between the adjusted values of 
$\zeta/R_\star\equiv n(a/R_\star)/\sqrt{1-b^2}$ and 
the square of the impact parameter, $b^2$ for various values of planetary radii
and limb darkening coefficients. Thin lines are for $p=0.01$, moderately
thick lines are for $p=0.1$ and thick lines are for $p=0.2$. The
continuous, long dashed, short dashed, dotted and dotted-dashed lines
represents the cases when limb 
darkening coefficients are $\gamma_1=\gamma_2=0$, $\gamma_1=\gamma_2=0.1$,
$\gamma_1=\gamma_2=0.2$, $\gamma_1=\gamma_2=0.3$
and $\gamma_1=\gamma_2=0.4$, respectively.}
\label{fig:corromb2}
\end{figure}

At this point we have
checked the correlations between an alternative parametrization
proposed by \citet{bakos2007}. In that work the light curve 
was parametrized by the equation
\begin{equation}
z^2(t)=\left(\frac{\zeta}{R_\star}\right)^2(1-b^2)(t-E)^2+b^2, \label{eq:uncorrtrparam}
\end{equation}
where $\zeta/R_\star=n(a/R_\star)/\sqrt{1-b^2}$. This parameter
is related to the duration of the transit, namely 
$(\zeta/R_\star)^{-1}=T_{\rm duration}/2$, where $T_{\rm duration}$
is the time between the instances when the center of the
planet crosses the limb of the star.
Since the above parametrization
of $z^2$ is linear in $b^2$, we have chosen $b^2$ instead of $b$ as 
an independent parameter. We have found that utilizing the parameter
set $\mathbf{a}'=(p,b^2,\zeta/R_\star,E)$ yields practically no correlation
between $b^2$ and $\zeta/R_\star$ for non-grazing eclipses and
increases only up to unity near $b\gtrsim 1-p$.
This correlation is plotted on Fig.~\ref{fig:corromb2} for various
planetary radii and limb darkening coefficients.

We should mention here that the recent work of \citet{carter2008}
gives an exhaustive analysis of the uncertainties and correlations
for various kind of transiting light curve characterizations. Their
work focuses on the analytical calculations for light curves with
no limb darkening and compares these results with
numerical derivations for the limb darkened cases.

\subsection{Uncertainties of the light curve parameters}
\label{subsec:uncertfilt}

\begin{figure}
\resizebox{8cm}{!}{\includegraphics{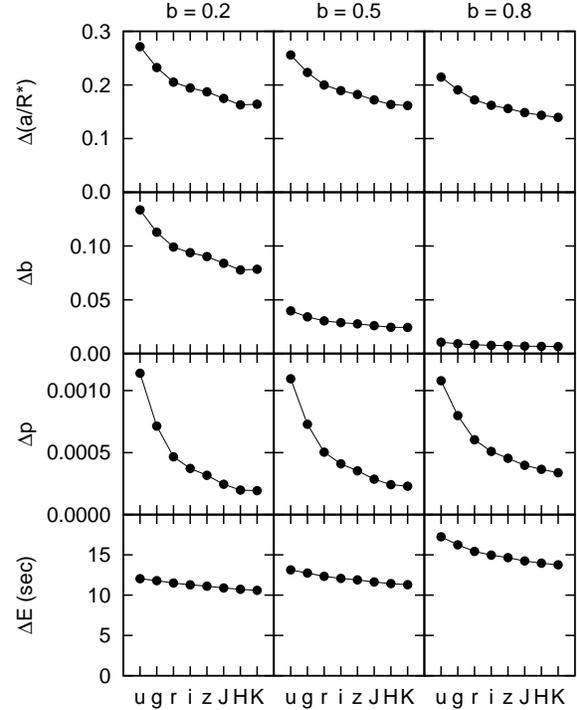}}
\caption{Uncertainties of the geometrical ratio $a/R_\star$, 
the impact parameter $b$, the fractional planetary radius 
$p\equiv R_{\rm p}/R_\star$
and the transit center time $E$ for a hypothetical transiting planet
with a radius of $p=0.1$ orbiting its Sun-like host star on an 
orbit with a semimajor axis of $a/R_\star=10$. The uncertainties
are plotted as the function of the photometric passbands, for
various impact parameters. See text for further details.}
\label{fig:uncertfilt}
\end{figure}

In this subsection we calculate the dependence of uncertainties of the light 
curve parameters $a/R_\star$, $E$ and $b$ 
(see equation~\ref{eq:classtrparam}) and the fractional planetary radius $p$ 
for various photometric passbands from near-ultraviolet to mid-infrared.
It is known that the limb-darkening parameters decrease for longer
wavelengths, therefore the transits themselves become shallower and
flattened. Using the Fisher information matrix method, as described earlier
gives a simple and straightforward way to obtain these uncertainties.
Assuming a solar-type star -- i.e. with metallicity of ${\rm [Fe/H]=0.00}$,
surface gravity of $\log g_\star=4.44$\,(CGS) and atmospheric temperature
of $T_{\rm eff}=5780\,{\rm K}$ -- we estimated these uncertainties
for photometric passbands $u'$, $g'$, $r'$, $i'$, $z'$, $J$, $H$ and $K$,
when such a star is transited by a hypotetical planet
with the orbital parameters of $a/R_\star=10$, $P=3.67\,{\rm days}$ 
and $p=R_{\rm p}/R_\star=0.1$. The appropriate limb darkening coefficients
for each filter have been obtained using the tables 
provided by \citet{claret2004}.
The results are plotted on Fig.~\ref{fig:uncertfilt} for various
impact parameters ($b=0.2$, $b=0.5$ and $b=0.8$).
During these estimations, the transits are assumed to be observed 
with a cadence of 10~seconds and with a photometric precision 
of $\Delta m=1.4\,{\rm mmag}$. Note that this photometric precision
is attainable by $\sim 1-1.2$\,m class telescopes 
for relatively bright stars ($z'\approx9.5\,{\rm mag}$) and using such cadence 
\citep[see e.g.][]{winn2007}.

It can clearly be seen from the plots of Fig.~\ref{fig:uncertfilt}
that the uncertainties for all of the parameters decrease 
for longer wavelengths; moreover, this decrement can reach a factor
of $\sim 10$ (between the near-ultraviolet and mid-infrared bands)
for the planetary radius. We note here that these analytical results
agrees well with numerical estimations (Joshua N. Winn, personal
communication).


\section{Discussion and summary}

In this paper the partial derivatives of the flux decrease function
of exoplanetary transits (or stellar binary eclipses)
has been calculated assuming a quadratic limb darkening law. 
These derivatives can then be applied for various analyses from which
we have demonstrated the correlation analysis of the 
limb darkening coefficients and two of the known transit 
light curve parameterizations. The most time consuming part
of the evaluation of the flux decrease (and its derivatives)
is the computing of the complete elliptic integrals.
Therefore, the calculation of the partial derivatives does not increase 
significantly the total computing time of both. 
The presented analytical analysis
of light curve parametrization is extremely fast comparing to
such an analysis based on Monte-Carlo methods: the integral 
in \eqref{eq:fisherintegral} should be calculated only once 
instead of the evaluation of the $\chi^2$ function $\mathcal{O}(10^4)$ 
times\footnote{which is necessary to obtain a 
reliable \emph{a posteriori} distribution of the parameters}. 
The knowledge of these derivatives also allows
straightforward calculations about how the variations in the 
orbital elements affect the light curves and the timings 
of the successive transits.

Here we note that such derivatives are also helpful in the implementation
of the Levenberg-Marquardt algorithm. Since this method requires the 
partial derivatives of the function to be adjusted, these must be evaluated
either analytically or numerically. The numerical evaluation of the 
partial derivatives requires the computation of the original function
twice in all directions of the parameter space. Therefore for
such a problem like transit light curve fitting, the numerical 
approximation requires approximately 10 times more computation time
(note that in practice the gain will be less, $\sim 8$ due to 
other overheads resulted by the computation of increased number of 
coefficients and the gain will clearly depend on the used programming
environment and its features).
Moreover -- because the derivatives of a transiting light curve function
lack the Lipschitz property of continuity -- the numerical approach
can also be unstable at the points of contacts. 

The routines for calculating both the transit decrease function
and its derivatives are 
available\footnote{http://szofi.elte.hu/\~{ }apal/utils/astro/ntiq/}
in Fortran77 and C languages along
with the codes used to calculate the correlations between the 
limb darkening parameters and the light curve parameters.

The recent review of \citet{southworth2008} concludes that 
the determination of some of the light curve parameters, 
especially the radius of the planet can be sensitive to the applied
limb darkening model and its parameters, yielding a possibility of 
systematic errors. Therefore, analytic description of transit light curves
for other different limb darkening laws would also be a point
of interest.

\section*{Acknowledgments}

The author would like to thank the hospitality and support of
the Harvard-Smithsonian Center for Astrophysics where this
work was partially carried out. The author acknowledges the support by 
the HATNet project, and NASA grant NNG04GN74G.
The author also thanks Eric Agol, G\'asp\'ar Bakos, 
Bence Kocsis and Joshua Winn for helpful comments and discussions. 


{}


\appendix

\section{Coefficients for the calculation of the flux decrease function}
\label{app:a}

In this section, the coefficients required to evaluate the 
flux decrease function -- as it is given by \eqref{eq:fluxdecrease} --
are summarized. The evaluation of the quantities $F_0$, $F_1$,
$F_K$, $F_E$, $F_{\Pi}$, $F_2$, $n$ and $k$ is done in two steps.
First, using Table~\ref{table:fdaux}, a set of
auxiliary variables should be evaluated, all of them are a function
only of $p$ and $z$, i.e. do not depend on the limb darkening coefficients.
Note that for a given geometrical case, not all of these quantities
have to be calculated, only those that referred to in the appropriate
row of Table~\ref{table:fdcoeffs}. (Moreover, it might happen that
for some of these equations the value of $p$ or $z$ are out of the
allowed domain if they are used in the cases when there is no need for them.)
Second, using Table~\ref{table:fdcoeffs}, the quantities $F_i$, $k$ and 
$n$ should by calculated by substituting the previously obtained
values of the auxiliary variables.

\begin{table*}
\caption{Auxiliary quantities for the calculation of the flux decrease.
Note that the quantities $a$, $b$, $k_0$ and $k_1$ are defined
similarly as were defined by \citet{mandel2002} and the former two
should not be confused with the same notations for the
semimajor axis and the impact parameter.}
\label{table:fdaux}
\begin{tabular}{llll}
$a=(p-z)^2$ &
$b=(p+z)^2$ \hspace*{5mm}
$t^2=p^2+z^2$ &
$\hat p=\sqrt{p(1-p)}$ &
$p'=\sqrt{1-p^2}$ \\
$C_I=\frac{2}{9\pi\sqrt{1-a}}$ & 
$C_{IK}=1-5z^2+p^2+ab$ & 
$C_{IE}=(z^2+7p^2-4)(1-a)$ & 
$C_{I\Pi}=-3\frac{p+z}{p-z}$ \\
$C_G=\frac{1}{9\pi\sqrt{pz}}$ &
$C_{GK}=3-6(1-p^2)^2-2pz(z^2+7p^2-4+5pz)$ &
$C_{GE}=4pz(z^2+7p^2-4)$ &
$C_{G\Pi}=-3\frac{p+z}{p-z}$ \\
\multicolumn{2}{l}{\ensuremath{T_I=\frac{2}{3\pi}\arccos(1-2p)-\frac{4}{9\pi}(3+2p-8p^2)\hat p}} &
$k_0=\arccos\left(\frac{p^2+z^2-1}{2pz}\right)$ &
$k_1=\arccos\left(\frac{z^2+1-p^2}{2z}\right)$ \\
\multicolumn{2}{l}{\ensuremath{G_0=\frac{p^2k_0+k_1-\sqrt{z^2-\frac{1}{4}(1+z^2-p^2)^2}}{\pi}}} &
\multicolumn{2}{l}{\ensuremath{G_2=\frac{k_1+p^2(p^2+2z^2)k_0-\frac{1}{4}(1+5p^2+z^2)\sqrt{(1-a)(b-1)}}{2\pi}}}
\end{tabular}
\end{table*}

\begin{table*}
\caption{Coefficients for 
the flux decrease function.}
\label{table:fdcoeffs}
\begin{tabular}{rlllllllll}
\hline
\hline
Step & Case	&
$F_0$ 	& 
$F_1$	&
$F_K$	&
$F_E$	&
$F_\Pi$	&
$F_2$	&
$k$	&
$n$	\\
\hline
1 & $\mathbf{A}$	&
$p^2$	&
$\frac{2}{3}(1-(p')^3)$	&
$0$	&
$0$	&
$0$	&
$\frac12p^4$	&
--	&
--	\\
2 & $\mathbf{A_G}$	&
$1$		&
$\frac23$		&
$0$		&
$0$		&
$0$		&
$\frac12$		&
--		&
--	\\
3 & $\mathbf{B}$	&
$p^2$			&
$\frac23$		&
$C_IC_{IK}$		&
$C_IC_{IE}$		&
$C_IC_{I\Pi}$		&
$\frac12p^2(p^2+2z^2)$	&
$\sqrt{\frac{4pz}{1-a}}$	&
$-\frac{4pz}{a}$	\\
4 & $\mathbf{B_T}$	&
$p^2$		&
$T_I$		&
$0$		&
$0$		&
$0$		&
$\frac12p^2(p^2+2z^2)$	&
--		&
--		\\
5 & $\mathbf{B_G}$	&
$G_0$		&
$\frac23$		&
$C_GC_{GK}$	&
$C_GC_{GE}$	&
$C_GC_{G\Pi}$	&
$G_2$		&
$\sqrt{\frac{1-a}{4pz}}$	&
$\frac{a-1}{a}$		\\
6 & $\mathbf{C}$	&
$p^2$		&
$\frac13$		&
$\frac{2}{9\pi}(1-4p^2)$	&
$\frac{8}{9\pi}(2p^2-1)$	&
$0$		&
$\frac32p^4$	&
$2p$		&
--		\\
7 & $\mathbf{C_T}$	&
$\frac14$		&
$\frac{1}{3}-\frac{4}{9\pi}$	&
$0$		&
$0$		&
$0$		&
$\frac{3}{32}$	&
--		&
--		\\
8 & $\mathbf{C_G}$	&
$G_0$		&
$\frac13$		&
$-\frac{1}{9\pi p}(1-4p^2)(3-8p^2)$	&
$\frac{1}{9\pi}16p(2p^2-1)$		&
$0$		&
$G_2$		&
$\frac{1}{2p}$	&
--		\\
9 & $\mathbf{D}$	&
$p^2$			&
$0$			&
$C_IC_{IK}$		&
$C_IC_{IE}$		&
$C_IC_{I\Pi}$		&
$\frac12p^2(p^2+2z^2)$	&
$\sqrt{\frac{4pz}{1-a}}$	&
$-\frac{4pz}{a}$	\\
10 & $\mathbf{E}$		&
$p^2$		&
$T_I$		&
$0$		&
$0$		&
$0$		&
$\frac12p^2(p^2+2z^2)$	&
--		&
--		\\
11 & $\mathbf{F}$	&
$G_0$		&
$0$		&
$C_GC_{GK}$	&
$C_GC_{GE}$	&
$C_GC_{G\Pi}$	&
$G_2$		&
$\sqrt{\frac{1-a}{4pz}}$	&
$\frac{a-1}{a}$		\\
12 & $\mathbf{G}$	&
$0$	&
$0$	&
$0$	&
$0$	&
$0$	&
$0$	&
--	&
--	\\
\hline
\hline
\end{tabular}
\end{table*}

\section{Parametric derivatives of the complete elliptic integrals}
\label{app:b}

The derivatives of the complete elliptic integrals of the
first and second kind are the following:
\begin{eqnarray}
\frac{\mathrm{d}K(k)}{\mathrm{d}k} & =  & \frac{E(k)}{k(1-k^2)}-\frac{K(k)}{k},\\
\frac{\mathrm{d}E(k)}{\mathrm{d}k} & = & \frac{E(k)-K(k)}{k}.
\end{eqnarray}
If the complete elliptic integral of the third kind is
defined with the sign convention as
\begin{equation}
\Pi(n,k)=\int\limits_0^{\pi/2}\frac{\mathrm{d}\varphi}{(1-n\sin^2\varphi)\sqrt{1-k^2\sin^2\varphi}}, \label{ellipticpidef}
\end{equation}
then its partial derivatives are the following:
\begin{eqnarray}
\frac{\partial\Pi(n,k)}{\partial n} & = &
\frac{1}{2(k^2-n)(n-1)}
\left[E(k)+\frac{(k^2-n)K(k)}{n}+\right. \nonumber \\
& & \left.+\frac{(n^2-k^2)\Pi(n,k)}{n}\right], \\
\frac{\partial\Pi(n,k)}{\partial k} & = &
\frac{k}{n-k^2}
\left[\frac{E(k)}{k^2-1}+\Pi(n,k)\right]. 
\end{eqnarray}
Throughout this paper we are using the sign convenction for $\Pi(n,k)$ as
it is defined by \eqref{ellipticpidef}.

\section{Coefficients for the the calculation of the partial derivatives
of the flux decrease function}
\label{app:c}

In this section, the coefficients required to evaluate the partial
derivatives of the flux decrease function are summarized,
which are needed to evaluate \eqref{eq:pdgfluxdecrease}. The evaluation
of these coefficients are also done in two steps.
First, using Table~\ref{table:fpartaux}, a set of
auxiliary variables should be evaluated, all of them are a function
only of $p$ and $z$ (i.e. do not depend on the limb darkening coefficients).
Note that for a given geometrical case, not all of these quantities
have to be calculated, only those that are referred to in the appropriate
row of Table~\ref{table:fpartcoeffs}. 
Second, using Table~\ref{table:fpartcoeffs}, the quantities $F_{i,g}$, 
$k_g$ and $n_g$ should by calculated (where $g$ is either $p$ or $z$),
by substituting the previously obtained values of the auxiliary variables.

\begin{table*}
\caption{Auxiliary quantities for the calculation of the 
partial derivatives of the flux decrease function.}
\label{table:fpartaux}
\begin{tabular}{ll}
$C_{IK,p}=+2p(1+2(p^2-z^2))$ &
$C_{IE,p}=14p(1-a)-2(p-z)(z^2+7p^2-4)$ \\
$C_{IK,z}=-2z(5+2(p^2-z^2))$ & 
$C_{IE,z}=2z(1-a)+2(p-z)(z^2+7p^2-4)$ \\
$C_{GK,p}=-2(p^2(12p+21z)+z(z^2-4)+2p(5z^2-6))$ &
$C_{GE,p}=4z(-4+21p^2+z^2)$ \\
$C_{GK,z}=2p(4-7p^2-10pz-3z^2)$ &
$C_{GE,z}=4p(-4+7p^2+3z^2)$ \\
$G_{0,p}=\frac{p}{\pi}\left(2k_0\right)$ & 
$G_{2,p}=\frac{p}{\pi}\left(2t^2k_0-4zp\sin k_0\right)$ \\
$G_{0,z}=\frac{p}{\pi}\left(-2\sin k_0\right)$ & 
$G_{2,z}=\frac{p}{\pi}\left(2zp k_0-(p^2+z^2+1)\sin k_0\right)$ \\
\end{tabular}
\end{table*}

\begin{table*}
\caption{Coefficients for partial 
derivatives of the flux decrease function.}
\label{table:fpartcoeffs}
\begin{tabular}{rlllllllll}
\hline
\hline
Step & Case	&
$\partial$	&
$F_{0,\partial}$ 	& 
$F_{1,\partial}$	&
$F_{K,\partial}$	&
$F_{E,\partial}$	&
$F_{2,\partial}$	&
$k_{\partial}$	&
$n_{\partial}$	\\
\hline
1 & $\mathbf{A}$	&
$p$	&
$2p$	&
$2pp'$	&
$0$	&
$0$	&
$2p^3$	&
--	&
--	\\
1 & $\mathbf{A}$	&
$z$	&
$0$	&
$0$	&
$0$	&
$0$	&
$0$	&
--	&
--	\\
2, 12 & $\mathbf{A_G}$, $\mathbf{G}$	&
$p$		&
$0$		&
$0$		&
$0$		&
$0$		&
$0$		&
--		&
--	\\
2, 12 & $\mathbf{A_G}$, $\mathbf{G}$	&
$z$		&
$0$		&
$0$		&
$0$		&
$0$		&
$0$		&
--		&
--	\\
3, 9 & $\mathbf{B}$, $\mathbf{D}$	&
$p$			&
$2p$			&
$0$			&
$C_IC_{IK,p}+C_IC_{IK}\frac{p-z}{1-a}$		&
$C_IC_{IE,p}+C_IC_{IE}\frac{p-z}{1-a}$		&
$2pt^2$	&
$\frac{2z(1+p^2-z^2)}{(1-a)^2k}$	&
$+\frac{4z(p+z)}{(p-z)a}$	\\
3, 9 & $\mathbf{B}$, $\mathbf{D}$	&
$z$		&
$0$			&
$0$			&
$C_IC_{IK,z}-C_IC_{IK}\frac{p-z}{1-a}$		&
$C_IC_{IE,z}-C_IC_{IE}\frac{p-z}{1-a}$		&
$2p^2z$	&
$\frac{2p(1-p^2+z^2)}{(1-a)^2k}$	&
$-\frac{4p(p+z)}{(p-z)a}$	\\
4, 10 & $\mathbf{B_T}$, $\mathbf{E}$	&
$p$		&
$2p$		&
$\frac{8p\hat p}{\pi}$		&
$0$		&
$0$		&
$2pt^2$	&
--		&
--		\\
4, 10 & $\mathbf{B_T}$, $\mathbf{E}$	&
$z$		&
$0$		&
$-\frac{8p\hat p}{3\pi}$		&
$0$		&
$0$		&
$2p^2z$	&
--		&
--		\\
5, 11 & $\mathbf{B_G}$, $\mathbf{F}$	&
$p$		&
$G_{0,p}$		&
$0$		&
$-\frac{F_K}{2p}-C_GC_{GK,p}$	&
$-\frac{F_E}{2p}+C_GC_{GE,p}$	&
$G_{2,p}$		&
$-\frac{1+p^2-z^2}{8kp^2z}$	&
$+\frac{2}{(p-z)^3}$		\\
5, 11 & $\mathbf{B_G}$, $\mathbf{F}$	&
$z$		&
$G_{0,z}$		&
$0$		&
$-\frac{F_K}{2z}-C_GC_{GK,z}$	&
$-\frac{F_E}{2z}+C_GC_{GE,z}$	&
$G_{2,z}$		&
$-\frac{1-p^2+z^2}{8kpz^2}$	&
$-\frac{2}{(p-z)^3}$		\\
6 & $\mathbf{C}$	&
$p$		&
$2p$		&
$0$		&
$\frac{4p}{9\pi}-\frac{1}{3\pi p}$	&
$\frac{28p}{9\pi}+\frac{1}{3\pi p}$	&
$4p^3$	&
$1$		&
--		\\
6 & $\mathbf{C}$	&
$z$		&
$0$		&
$0$		&
$-\frac{20p}{9\pi}+\frac{1}{3\pi p}$	&
$\frac{4p}{9\pi}-\frac{1}{3\pi p}$	&
$2p^3$		&
$1$		&
--		\\
7 & $\mathbf{C_T}$	&
$p$		&
$1$		&
$\frac{2}{\pi}$	&
$0$		&
$0$		&
$\frac12$		&
--		&
--		\\
7 & $\mathbf{C_T}$	&
$z$		&
$0$		&
$-\frac{2}{3\pi}$	&
$0$		&
$0$		&
$\frac14$	&
--		&
--		\\
8 & $\mathbf{C_G}$	&
$p$		&
$G_{0,p}$	&
$0$		&
$\frac{3+16p^2(2-9p^2)}{18\pi p^2}$	&
$\frac{72p^2-2}{9\pi}$	&
$G_{2,p}$		&
$-\frac{1}{4p^2}$	&
--		\\
8 & $\mathbf{C_G}$	&
$z$		&
$G_{0,z}$	&
$0$		&
$\frac{3+ 8p^2(1-6p^2)}{18\pi p^2}$	&
$\frac{24p^2-14}{9\pi}$	&
$G_{2,z}$		&
$-\frac{1}{4p^2}$	&
--		\\
\hline
\hline
\end{tabular}
\end{table*}

\label{lastpage}

\end{document}